# UNIFIED EXTERNAL STAKEHOLDER ENGAGEMENT AND REQUIREMENTS STRATEGY

Ahmed Abdulaziz Alnhari[1] and Rizwan Qureshi [2]

Department of Information Technology, Faculty of Computing and Information Technology, King Abdul-Aziz University, Jeddah 80213, Saudi Arabia

*ABSTRACT*

*Understanding stakeholder needs is essential for project success, as stakeholder importance varies across projects. This study proposes a framework for early stakeholder identification and continuous engagement throughout the project lifecycle. The framework addresses common organizational failures in stakeholder communication that lead to project delays and cancellations. By classifying stakeholders by influence and interest, establishing clear communication channels, and implementing regular feedback loops, the framework ensures effective stakeholder involvement. This approach allows for necessary project adjustments and builds long-term relationships, validated by a survey of IT professionals. Engaging stakeholders strategically at all stages minimizes misunderstandings and project risks, contributing to better project management and lifecycle outcomes.*

## 1. INTRODUCTION

This paper examines the critical role of stakeholder analysis and management in project development, emphasizing the need to understand stakeholder needs and requirements [1]. Early identification and engagement of stakeholders help create a shared vision, which is essential for defining project scope, reducing ambiguities, and planning effectively for costs and timelines [2-3]. One of the core objectives of this paper is to explore and evaluate the various stakeholder management techniques that have been employed across different industries and fields over the years. This exploration is aimed at understanding the diversity of approaches and identifying those that are most effective for different types of projects, irrespective of their industry [4]. By analysing these techniques, the paper seeks to offer insights into the best practices that can be adapted and applied to a wide range of projects. This inclusive approach acknowledges that stakeholder management is not a one-size-fits-all process; rather, it varies significantly across sectors due to differing stakeholder expectations, regulatory environments, and project goals [5]. Thus, the paper aspires to provide a comprehensive framework that can be adapted to suit diverse project needs, ensuring that stakeholder management strategies are both flexible and effective.

In addition to examining stakeholder management techniques, the paper places a strong emphasis on the classification of stakeholders [6]. Recognizing that stakeholders are not a homogenous group, the paper advocates for a systematic categorization of stakeholders into distinct groups based on their characteristics and relationships to the project. Examples of these groups include Business-to-Business (B2B) stakeholders, who are often partners or suppliers with direct business interests, and Small and Medium Businesses (SMBs), who may have different concerns and levels of influence [7]. By categorizing stakeholders into such groups, the paper aims to provide a more nuanced understanding of the types of stakeholders involved in various projects. This classification process helps in identifying common patterns of stakeholder behaviour and expectations, which in turn facilitates the development of customized strategies for engaging each group effectively [8].

The importance of stakeholder engagement is a recurring theme throughout the paper, emphasizing that engagement should be viewed from both the stakeholders' and the organization's perspectives [9]. From the stakeholders' perspective, effective engagement ensures that their voices are heard, their concerns are addressed, and their interests are considered, leading to a sense of ownership and commitment to the project. From the organization's perspective, engaging stakeholders effectively helps in maintaining a smooth project timeline, minimizing conflicts, and ensuring that stakeholders' contributions are meaningful and aligned with project goals [10]. The paper argues that such dual-focused engagement is critical for the success of any project, as it not only enhances stakeholder satisfaction but also strengthens the project's overall outcomes.

The paper is structured to first review existing literature on stakeholder analysis, identifying gaps in current practices (Section 2). It then defines the specific challenges in stakeholder management (Section 3), followed by presenting a novel solution that integrates early identification, classification, and tailored engagement strategies (Section 4).

## 2. RELATED WORK

Brugha et al. [1] emphasized the significance of stakeholder analysis and outlined key factors to consider before conducting such analysis. These factors include the purpose, time dimensions, frame, and context. However, it should be noted that the method proposed may primarily be applicable to large B2B organizations. For smaller organizations that regularly interact with a wide range of stakeholders, the applicability of this method may be limited. In contrast, Alsulaimi et al. [2] presented an alternative approach to stakeholder analysis using a mobile application. This solution focuses on two key aspects: demographic factors (such as age, sex, and education level) and project duration. By considering these factors, the proposed solution aims to define stakeholders, enhance communication effectiveness, and provide project managers with tools for planning and managing IT projects. However, the suggested solutions face challenges, including resistance to data sharing, difficulties in addressing diverse stakeholder needs and audiences, varying requirements among stakeholders, and limited resources or communication tools.

Sapapthai et al. [3] stressed the importance of stakeholders in identifying the primary parties associated with a project. According to Sasanee et al., analyzing stakeholders is crucial for understanding the complex relationships among all involved parties. The study categorizes stakeholder analysis into three primary categories: (1) identification of stakeholders, (2) differentiation and categorization of stakeholders, and (3) investigation of relationships among stakeholders. However, the field evidence supporting these concepts is somewhat lacking, and the discussion often covers a broad range of topics without providing clear specifics. Yang [4] described how analyzing the decision-making group should be a priority. Rebecca's study highlighted two perspectives for stakeholder analysis: empiricism and rationalism, emphasizing that no single method is perfect. Instead, choosing an analytical perspective involves careful consideration of 'when, what, and how' to achieve project objectives. It is surprising that the study does not advocate for a particular solution or method to develop a suitable approach. The methods of analyzing stakeholders, understanding their relationship to the project, and managing communication between them are essential for the project's continuation.

Table 1.  The main limitations of related work.

| Title | Limitations |
|---|---|
| Stakeholder analysis [1] | It should be noted that the method proposed may primarily apply to big B2B organizations, for small organizations who regularly interact with a wide range of stakeholders, the applicability of this method may be limited. |
| Management of Stakeholder Communications in IT Projects [2] | The suggested solutions also face challenges, including resistance to data sharing, difficulties in catering to diverse stakeholder needs and audiences, varying requirements among stakeholders, and limited resources or communication tools. |
| A Stakeholder Analysis Approach for Area Business Continuity Management: A Systematic Paper [3] | The evidence supporting these concepts in the field is somewhat lacking, and the text generally covers a broad range of topics without providing clear specifics. |
| An investigation of stakeholder analysis in urban development projects: Empirical or rationalistic perspectives [4] | The methods of analyzing stakeholders, their relation to the project, and the importance of managing the communication between them are extraordinary for the project to continue [1-4]. |
| Stakeholder analysis and engagement in projects: From stakeholder relational perspective to stakeholder relational ontology [5] | The approach was limited to one case study, and the project actors did not actually utilize the conceptual approach based on Actor-Network Theory (ANT) in practice, which limits the validation and practical application of the framework. |
| Stakeholder Analysis and Their Attitude towards PPP Success [6] | The research should consider specific country circumstances, PPP forms, contracts, and traditions to adopt a more stakeholder-oriented perspective in project management. |
| Stakeholder analysis in projects: Challenges in using current guidelines in the real world [7] | The author acknowledges the time-consuming nature of thorough stakeholder analysis, and the dilemma project managers may face between conducting a comprehensive analysis and the need to initiate the project promptly. |
| Stakeholder Analysis for Smart City Development Project: An Extensive Literature Paper [8] | The proposed solutions focused on smart city development projects more than the main purpose solution of the research which is stakeholder analysis. |
| StakeSoNet: Analysis of Stakeholders Using Social Networks [9] | The scalability of the proposed method for large-scale software projects may pose technical and computational challenges. |
| Project Stakeholders: Analysis and Management Processes [10] | The difficulty of adapting the stakeholder analysis approach to different types and sizes of projects. |

Missonier et al. [5] addressed the limitations of previous studies in stakeholder analysis and proposed a conceptual approach to overcome these challenges. Their approach involves shifting from a stakeholder relational perspective based on Social Network Theory to a stakeholder relational ontology using Actor-Network Theory. Their findings highlight the dynamic and evolving nature of stakeholder relationships. However, the approach was tested in only one case study, and project actors did not actively utilize the Actor-Network Theory in practice, which limits the validation and practical application of the framework. In a return to basic stakeholder analysis, Wegrzyn et al. [6] used a benefit-engagement model based on attributes related to sustainable development and engagement, including time and scope perspectives. Unfortunately, their research suggests that specific country circumstances, PPP forms, contracts, and traditions should be considered to adopt a more stakeholder-oriented perspective in project management.

The study by Jepsen et al. [7] highlighted that identifying stakeholders is more crucial than merely analyzing them. The study suggests approaching stakeholder analysis as an ongoing learning process, engaging in dialogue with stakeholders early on to consider their input. However, the authors acknowledge the time-consuming nature of thorough stakeholder analysis and the dilemma project managers face between conducting comprehensive analysis and the need to start projects promptly. Jayasena et al. [8] underscored the paramount importance of timely and effective stakeholder consultation for project success. They proposed a framework for enhancing stakeholder engagement based on a review of 31 papers. This framework involves identifying and categorizing stakeholders, both internal and external, and serves as a basis for effective engagement. Although the proposed solutions were focused on smart city development projects, they provide valuable insights into stakeholder analysis more broadly.

Stakeholders play a key role during the Software Development Life Cycle (SDLC). Hassan et al. [9] discussed stakeholder analysis in the context of an Intelligent Energy System (IES) project, highlighting the lack of stakeholder participation during requirements elicitation. Their proposed solution was a method for stakeholder analysis using social networks, employing the CRM-78-APGMIR algorithm to select stakeholders based on confidence values. However, the scalability of this method for large-scale software projects may pose technical and computational challenges. In discussing stakeholder importance, Riahi [10] emphasized that stakeholders' opinions should be considered. Riahi advocated for a stakeholder management plan that defines and understands stakeholders as individuals, groups, or organizations that can influence or be influenced by the project. Nevertheless, the difficulty of adapting stakeholder analysis approaches to different types and sizes of projects remains a significant concern.

The study in [11] addresses the challenges of sustaining project management practices within virtual team environments. It highlights the difficulties of engaging stakeholders effectively and managing knowledge efficiently when teams operate remotely, leading to communication barriers, reduced team cohesion, and potential project inefficiencies. To address these issues, the authors propose a sophisticated framework that integrates advanced knowledge management techniques with proactive stakeholder engagement strategies, implemented through digital tools. These tools are designed to enhance communication and collaboration, ensuring effective knowledge sharing while aligning project goals with stakeholder expectations. However, the framework's reliance on significant technological integration and the need for high digital literacy among stakeholders could potentially exclude less tech-savvy participants and place additional resource demands on smaller organizations or projects with limited budgets.

A study conducted in [12] seeks to address the complex challenge of integrating diverse stakeholder groups effectively. Recognizing the substantial influence these groups have on project outcomes; a comprehensive stakeholder management framework is introduced at enhancing engagement from the project's outset. This framework focuses on systematically identifying, categorizing, and continuously engaging stakeholders, incorporating tools to map their influences and expectations. The strategy is designed to align project strategies closely with stakeholder needs, maximizing value creation. However, the framework faces significant challenges due to the regulatory complexities and high public stakes typically associated with projects such as hospital developments. These factors necessitate extensive resource investment and constant reassessment of stakeholder dynamics, posing managerial challenges, especially under tight schedules and budget constraints.

Klaus-Rosińska and Iwko [13] focused their research on the crucial role of stakeholder management in promoting sustainable project management within small construction companies. They pointed out that many small firms often underestimate the strategic importance of effective stakeholder engagement, which can lead to challenges in achieving sustainability and project success. To address this gap, they proposed a specialized stakeholder management framework tailored to the unique needs and constraints of small construction businesses. This framework aims to improve stakeholder interactions and ensure that their needs and goals align with the project's sustainability objectives. However, despite its potential benefits, implementing the framework poses substantial challenges. Small companies frequently lack the necessary resources and expertise to adopt complex management strategies and may be reluctant to change established practices due to entrenched cultural norms and organizational structures, hindering the effective adoption of the proposed solutions.

The paper in [14] explores the integration of various processes, methods, tools, and techniques from management science into project management practices. Its objective is to evaluate how these elements can be effectively combined to improve project management across different industries. The paper identifies a gap in existing project management approaches, noting the absence of a unified framework that merges advanced management science techniques with traditional project management tools. To bridge this gap, Doe and Smith propose an innovative, holistic framework that leverages cutting-edge management theories alongside proven practical tools to enhance project outcomes. However, they acknowledge that the complexity and theoretical nature of their proposed methods may pose challenges for adoption, particularly in sectors where hands-on, practical project management styles are more prevalent.

In a related discussion, a study [15] highlights the need for more comprehensive research into stakeholder management strategies that are adaptable to today's dynamic, global project environments. The authors call for the exploration of new, adaptive strategies as essential for advancing the field. However, their approach relies heavily on the academic community to address these gaps without offering concrete, immediate solutions to pressing stakeholder management challenges. While their call underscores the urgency for innovative research, it also highlights the potential slow pace of translating academic findings into practical, actionable strategies.

A brief summary of limitations in the related work is shown in Table 1.

## 3. PROBLEM DEFINITION

Klaus-Rosińska and Iwko [14] explore the critical role of stakeholder management in enhancing sustainable project management within small construction companies. They identify a common issue in these firms: the underestimation of effective stakeholder engagement, which is essential for project success and sustainability. This oversight often leads to inefficiencies and difficulties in meeting project objectives, particularly in small-scale operations that may lack robust processes. To address this issue, the authors propose a specialized stakeholder management framework tailored to the unique dynamics of small construction firms. This framework aims to facilitate better engagement by systematically identifying, analysing, and involving stakeholders in ways that align with the project's sustainability goals. The goal is to ensure that all stakeholder interests are considered and integrated throughout the project lifecycle to enhance outcomes and create value. However, implementing such a framework presents challenges. The primary concern is its practical application in the context of small businesses, which often face resource constraints and may lack the expertise required for sophisticated stakeholder management practices. Additionally, cultural resistance within these firms to adopting new methodologies that differ from their established practices could hinder the effectiveness of the framework. This resistance could limit the framework's potential to bring about meaningful change in stakeholder management and, consequently, in the sustainability of project management practices within small construction companies.

Following is a research question based on the problem discussed in [14].

- How can small construction firms effectively adopt stakeholder management practices to boost project sustainability and success, while navigating challenges like limited resources and resistance to change?

This question seeks to explore the methods and strategies that could be adapted to the specific conditions of small construction firms to improve their stakeholder management practices. It acknowledges the challenges these companies face, such as resource constraints and cultural resistance to change, and aims to identify practical, scalable solutions that could be readily integrated into their existing project management frameworks to achieve better sustainability and project outcomes.

## 4. THE PROPOSED SOLUTION

Small construction companies can effectively implement stakeholder management practices to enhance project sustainability and success by adopting several key strategies. Prioritizing clear, consistent, and open communication is essential. Establishing regular communication channels for updates and feedback ensures that stakeholders are informed about project progress, changes, and challenges, which helps build trust and manage expectations. Early identification and understanding of stakeholders' interests, concerns, and expectations allow companies to tailor their engagement strategies effectively. This early engagement ensures that stakeholders feel valued and involved, which can reduce resistance and align their goals with the project's objectives. To thrive in managing complex projects, small construction companies must take a refined approach to stakeholder management. This begins with the initial step of creating a comprehensive stakeholder assessment form tailored to each project's unique scale and requirements. This form is not merely an administrative tool but a strategic one, capturing essential data about stakeholders' roles, expectations, and potential influence on project outcomes. Properly gathering and analysing this information allows project managers to map out the landscape of stakeholder interests and prioritize them accordingly, ensuring that critical relationships are nurtured and managed effectively.

Leveraging affordable technology, such as project management software and collaborative platforms, can significantly improve efficiency. These tools streamline communication, reporting, and collaboration, making stakeholder management more manageable even with limited resources. Additionally, fostering a collaborative culture both within the company and with external stakeholders can help minimize resistance to new methodologies. Involving stakeholders in decision-making processes and demonstrating how their input impacts project success fosters buy-in and support. Building on the foundation of thorough data collection and analysis, developing a stakeholder engagement strategy is the next critical step. This strategy must be robust and flexible, outlining detailed methods for continuous and meaningful interactions with all stakeholders. It should cater to varying stakeholder profiles, incorporating strategies that range from direct meetings and digital communication to stakeholder workshops and regular project updates. The engagement strategy should also specify the frequency and format of interactions, ensuring that stakeholders are kept informed and can provide ongoing feedback that could influence the project's direction.

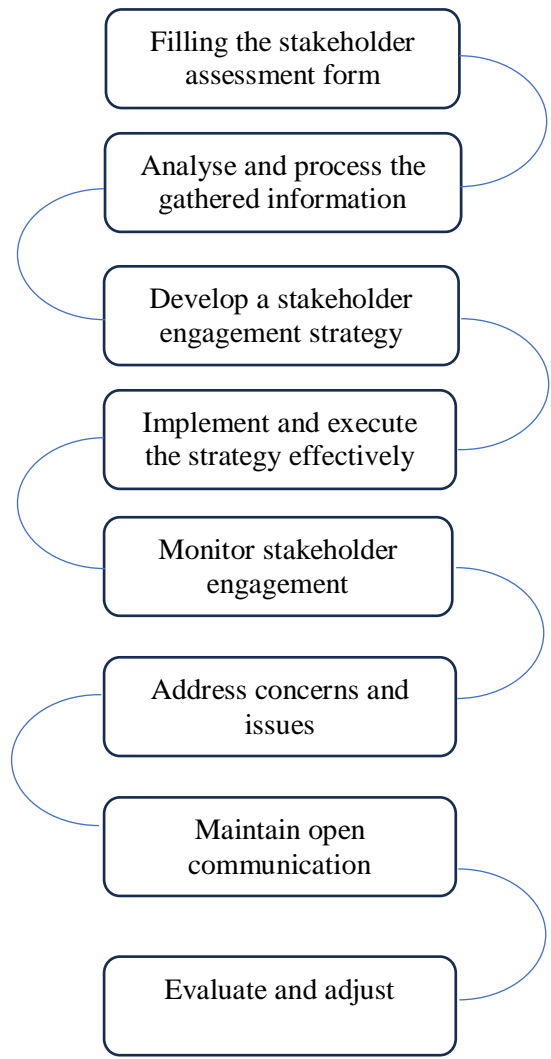

Figure 1. The graphic presentation of the proposed framework

Training and educating employees on stakeholder management principles, such as effective communication, negotiation, and conflict resolution, is another crucial step. By equipping staff with the necessary skills, small construction firms can handle stakeholder relationships more effectively. Adapting and scaling best practices to suit the company's size and context is also vital. Instead of adopting complex frameworks designed for larger organizations, small companies should customize practices that align with their specific needs and resources. Implementing the engagement strategy requires meticulous execution, where project managers and their teams are clear on their roles and responsibilities. Modern project management tools can significantly aid this process, helping to schedule, track, and report on engagement activities, ensuring that nothing is overlooked and that all actions align with the project's strategic goals.

However, merely implementing a strategy is not enough. Ongoing monitoring and evaluation of the engagement process are crucial. This includes establishing key performance indicators for stakeholder engagement and using these metrics to continually assess the effectiveness of interaction efforts. Regular monitoring not only helps in adjusting strategies to better meet stakeholder needs but also ensures that the project can adapt to any new challenges or opportunities that arise.

Addressing concerns promptly and maintaining open communication channels are paramount. This involves setting up responsive mechanisms where stakeholders can voice their concerns or provide suggestions. Regular feedback sessions and updates help maintain transparency and trust, fostering a collaborative project environment. Finally, the iterative evaluation of stakeholder engagement practices is essential. This step involves periodically revisiting the engagement strategy and making adjustments based on stakeholder feedback and the evolving project context. Such a dynamic approach ensures that the project remains responsive to stakeholders' needs and agile enough to adapt to changes, thereby improving the chances of project success and long-term sustainability. By focusing on these strategies, small construction companies can effectively manage their stakeholders, ensuring not only the successful completion of projects but also their sustainability. This approach will help build lasting relationships with stakeholders, leading to future business opportunities and growth. Embracing a comprehensive and structured approach to stakeholder management allows small construction companies to enhance their project management capabilities, leading to better project outcomes and stronger relationships with all project stakeholders. This proactive and strategic approach not only addresses immediate project challenges but also builds a resilient framework for managing future projects and stakeholder relationships more effectively. The graphic representation of the proposed solution is shown in figure 1.

### 4.1. Steps to Implement the Proposed Framework

To provide more details on how the proposed framework for stakeholder engagement can be implemented in real-world scenarios, the following steps can be incorporated.

#### 4.1.1. Initial Stakeholder Identification and Assessment

Begin by using a structured stakeholder assessment form to gather essential data about potential stakeholders. This data should include their roles, interests, influence, and expectations related to the project. For example, in a software development project, stakeholders might include customers, end-users, developers, testers, project managers, and business analysts. Each of these groups will have different priorities and levels of influence.

#### 4.1.2. Categorizing Stakeholders

Once identified, stakeholders should be categorized based on their influence and interest. High-influence, high-interest stakeholders, such as senior management or key clients, should receive more frequent updates and involvement in decision-making processes. In contrast, low-influence, low-interest stakeholders might only need periodic updates. Tools like power-interest grids can help visualize and prioritize stakeholder engagement efforts.

#### 4.1.3. Developing a Communication Plan

A tailored communication plan should be established, outlining how and when to communicate with different stakeholder groups. This plan should specify the channels (e.g., emails, meetings, newsletters), frequency, and content of communications. For example, key stakeholders might be engaged through monthly strategic meetings, while others receive quarterly newsletters.

### 4.1.4. Ongoing Engagement and Feedback Mechanisms

Continuous engagement is critical. Regular check-ins, surveys, and feedback loops can help capture stakeholder input throughout the project. This can be implemented through scheduled meetings, feedback forms, or even informal check-ins. For instance, during the development of a new product, stakeholders could be involved in user testing phases, providing feedback that can be immediately acted upon.

### 4.1.5. Documentation and Alignment

All interactions, agreements, and feedback should be thoroughly documented. This documentation serves as a reference to track stakeholder expectations and project progress, ensuring alignment. For example, meeting minutes, decision logs, and action items should be recorded and shared with relevant stakeholders to maintain transparency and accountability.

### 4.1.6. Conflict Resolution and Negotiation

Establishing a clear process for handling conflicts and disagreements is vital. This might include setting up a conflict resolution committee or defining escalation paths. For instance, if a conflict arises between the development team and a stakeholder over feature priorities, a facilitated negotiation session could be conducted to align both parties on a common goal.

### 4.1.7. Final Review and Evaluation

Towards the end of the project, conduct a final review with stakeholders to assess whether their needs and expectations have been met. This review should involve evaluating the effectiveness of the engagement strategies and gathering feedback for future projects. This can be implemented through project debrief meetings, post-project surveys, or focus group discussions.

### 4.1.8. Training and Capacity Building

To ensure the effective implementation of this framework, project teams should be trained in stakeholder management techniques. This includes communication skills, conflict resolution, and negotiation. Workshops, training sessions, and access to stakeholder management tools can enhance the team's ability to engage stakeholders effectively.

## 5. VALIDATION OF THE PROPOSED SOLUTION

To effectively validate the proposed solution for enhancing stakeholder engagement and requirements identification, a detailed analysis is structured around four specific goals, aligning with data collected and presented in various tables and diagrams from the related work. The proposed solution aims to improve stakeholder engagement and requirements identification in projects by focusing on four critical goals. Each goal targets specific aspects of stakeholder management, ensuring comprehensive coverage and alignment with best practices in project management.

- Goal 1: Early Identification of Stakeholders Before Project Initiation

Objective: The primary objective of this goal is to identify stakeholders at the earliest stage of the project lifecycle. Early identification allows for a clear understanding of each stakeholder's desires, expectations, and potential impact throughout the project.

Importance: This early engagement is crucial as it establishes the foundation for effective communication and stakeholder management, ensuring that all relevant voices are considered from the planning phase onward.

- Goal 2: Alignment of Stakeholder Needs with Organizational Capabilities

Objective: Once stakeholders are identified, this goal focuses on aligning their needs and requirements with the organization's capabilities and project scope.

Importance: Alignment is essential to ensure that the project can meet stakeholder expectations without overextending resources. It involves assessing the organization's ability to deliver on stakeholder demands, which is critical for maintaining credibility and trust.

- Goal 3: Facilitation of Effective Communication and Negotiation

Objective: After identifying stakeholders and aligning their needs with the project's capabilities, the next goal is to establish and maintain open and effective communication channels.

Importance: This goal facilitates ongoing dialogue and negotiation, vital for addressing any concerns that arise during the project. Effective communication keeps stakeholders engaged and committed to the project and helps manage and adjust expectations as the project evolves.

- Goal 4: Thorough Documentation and Risk Management

Objective: The final goal emphasizes the importance of thorough documentation and proactive risk management.

Importance: Documenting interactions, agreements, and plans with stakeholders maintains a clear record that can be referenced throughout the project. This documentation is crucial for managing risks associated with miscommunications or misunderstandings. It also serves as a vital tool for conflict resolution and ensures that all parties are aware of their roles, responsibilities, and the project's progress.

Together, these four goals provide a structured approach to stakeholder management that addresses immediate project needs and contributes to the long-term success and sustainability of project outcomes. By systematically addressing each of these areas, projects are more likely to meet or exceed stakeholder expectations and achieve their intended objectives.

## 5.1. Goal 1 - Early Identification of Stakeholders before Project Initiation

The primary objective of Goal 1 is to ensure stakeholders are identified early in the project to understand their desires and expectations throughout the project lifecycle. The validation of this goal is supported by data from Table 2, showing a high percentage (48.00%) of responses rating the effectiveness of stakeholder identification as "very high." This suggests that early identification of stakeholders can significantly influence project outcomes by aligning expectations from the outset. However, the varied responses across categories also indicate a need for a more standardized approach to capturing stakeholder information consistently.

International Journal of Software Engineering & Applications (IJSEA), Vol.15, No.4, September 2024

Table 2. Cumulative analysis of Goal 1

| Q. No. | Very low | Low | Nominal | High | Very high |
|---|---|---|---|---|---|
| Q1 | 2 | 1 | 3 | 4 | 2 |
| Q2 | 0 | 12 | 4 | 4 | 6 |
| Q3 | 9 | 15 | 14 | 18 | 12 |
| Q4 | 16 | 12 | 24 | 12 | 28 |
| Q5 | 70 | 40 | 55 | 40 | 35 |
| Avg. | 2.40% | 5.20% | 27% | 18.40% | 48.00% |

## 5.2. Goal 2 - Alignment of Stakeholder Needs with Organizational Capabilities

Goal 2 focuses on ensuring that stakeholders' needs and requirements align with the organization's capacity to deliver on these expectations. Data from Table 3 reflects substantial acknowledgment (49% "very high") of the effectiveness of organizational assessments post-stakeholder identification. This implies that most projects succeed in aligning stakeholder needs with organizational capabilities, though there remains room for improvement in formalizing these assessments to ensure uniformity and avoid discrepancies in resource allocation.

Table 3. Cumulative analysis of Goal 2

| Q. No. | Very low | Low | Nominal | High | Very high |
|---|---|---|---|---|---|
| Q1 | 3 | 2 | 14 | 1 | 4 |
| Q2 | 2 | 6 | 13 | 2 | 4 |
| Q3 | 18 | 9 | 21 | 21 | 6 |
| Q4 | 20 | 20 | 12 | 20 | 20 |
| Q5 | 45 | 55 | 40 | 50 | 55 |
| Avg. | 3% | 4% | 26% | 18% | 49% |

Table 4. Cumulative analysis of Goal 3

| Q. No. | Very low | Low | Nominal | High | Very high |
|---|---|---|---|---|---|
| Q1 | 6 | 6 | 0 | 28 | 65 |
| Q2 | 2 | 2 | 9 | 24 | 55 |
| Q3 | 4 | 4 | 6 | 16 | 65 |
| Q4 | 8 | 8 | 6 | 16 | 65 |
| Q5 | 2 | 2 | 3 | 24 | 70 |
| Q6 | 6 | 6 | 12 | 24 | 50 |
| Avg. | 5% | 5% | 3% | 22% | 65% |

## 5.3. Goal 3 - Facilitation of Communication and Negotiation

After establishing initial alignment, Goal 3 aims to facilitate ongoing communication between project parties to deepen understanding and negotiate any differences. The data from Table 4, with 62% "very high" ratings, indicates strong performance in this area, suggesting that structured communication pathways effectively enhance stakeholder relations and project clarity. However, continued evaluation of these interactions is crucial to maintain this effectiveness throughout the project duration.

## 5.4. Goal 4 - Documentation and Risk Management

The fourth goal addresses the need for thorough documentation and proactive risk management to mitigate potential failures. While specific data on this goal wasn't detailed, the reference to high demand for "paper steps" implies recognition of the importance of documentation in safeguarding project interests and ensuring all phases are clearly articulated and agreed upon by all stakeholders. Table 5 shows that 56% of respondents support Goal 4 very highly.

Table 5. Cumulative analysis of Goal 4

| Q. No. | Very low | Low | Nominal | High | Very high |
|---|---|---|---|---|---|
| Q1 | 1 | 4 | 12 | 16 | 65 |
| Q2 | 4 | 2 | 12 | 16 | 50 |
| Q3 | 4 | 4 | 21 | 12 | 45 |
| Q4 | 4 | 0 | 9 | 8 | 65 |
| Avg. | 2% | 3% | 26% | 13% | 56% |

## 5.5. Final Cumulative Analysis of Four Goals

The validation of the proposed solution through these four goals demonstrates a strong foundational approach to stakeholder management. However, the data also highlights areas for improvement, particularly in standardizing processes and ensuring consistent application across all project types and sizes. Further enhancements could include developing more robust training modules for project managers on stakeholder engagement, integrating advanced digital tools to streamline data collection and analysis, and establishing more rigorous follow-up mechanisms to adapt strategies based on real-time feedback and evolving project conditions. This structured approach to validating the proposed solution not only confirms its potential effectiveness but also outlines clear pathways for refining the strategy to better meet the nuanced needs of various stakeholders and project environments. As shown in Table 6, the cumulative analysis reveals that 3% of professionals report a very low effect of the proposed solution, 4% report a low effect, 20% report a nominal effect, 17% highly support Goals 1 through 4, and 54% very highly favour the proposed solution. Figure 2 graphically represents the final cumulative analysis of the four goals, clearly supporting the validation of the proposed solution.

Table 6. Final Cumulative Analysis of Four Gaols

| Goal. No. | Very low | Low | Nominal | High | Very high |
|---|---|---|---|---|---|
| Goal 1 | 2% | 5% | 27% | 18% | 48% |
| Goal 2 | 3% | 4% | 26% | 18% | 49% |
| Goal 3 | 5% | 5% | 3% | 22% | 65% |
| Goal 4 | 2% | 3% | 26% | 13% | 56% |
| Total | 12% | 17% | 82% | 71% | 218% |
| Avg. | 3% | 4% | 20% | 17% | 54% |

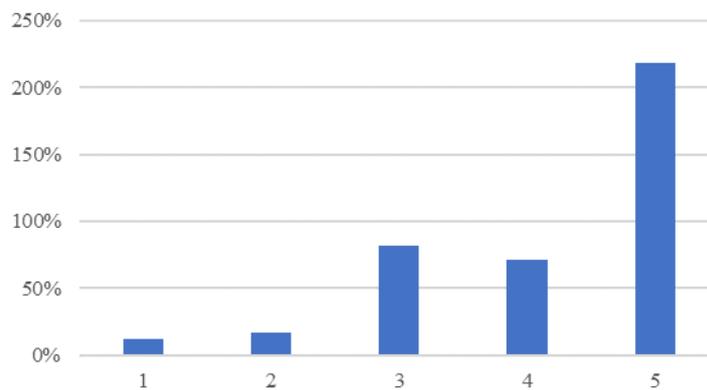

Figure 2. Final cumulative analysis of four goals

## 6. COMPARISON OF THE PROPOSED FRAMEWORK WITH EXISTING FRAMEWORKS

To provide a comprehensive understanding of the proposed framework's unique contributions, it is essential to compare it with existing frameworks and best practices. This comparison will help contextualize the research and highlight the specific advantages of the proposed approach.

- Traditional Stakeholder Analysis Approaches: Traditional stakeholder analysis frameworks, like those proposed by Brugha et al. [1], focus on factors such as the purpose, time dimensions, frame, and context of the stakeholder analysis. These methods are often suited to large B2B organizations where stakeholder relationships are more stable and predictable. However, these approaches can be less effective for smaller organizations or projects that interact with a more diverse and dynamic range of stakeholders.

  o Unique Contribution: The proposed framework distinguishes itself by advocating for continuous stakeholder engagement rather than a one-time analysis. This ongoing interaction helps accommodate the needs of diverse stakeholders, making it more adaptable to projects of varying sizes and industries. By using a dynamic categorization system based on influence and interest, the framework allows for real-time adjustments to stakeholder strategies, enhancing flexibility and relevance.

    - Technology-Driven Solutions: Alsulaimi et al. [2] explored the use of mobile applications for stakeholder analysis, focusing on demographic factors and project duration to define stakeholders and enhance communication. While innovative, these technology-driven approaches often face challenges such as resistance to data sharing and the complexity of addressing diverse stakeholder needs.

      o Unique Contribution: The proposed framework not only leverages technology for communication but also emphasizes the human element in stakeholder engagement. It incorporates regular feedback loops and open communication channels, ensuring that technology enhances rather than replaces meaningful stakeholder interactions. This balanced approach mitigates the risk of technology resistance and ensures stakeholder buy-in.

    - Actor-Network Theory (ANT) and Relational Approaches: Missonier et al. [5] suggested using Actor-Network Theory to understand the dynamic nature of stakeholder

relationships. While ANT offers valuable insights into the evolving nature of these relationships, its practical application can be limited by its complexity and the requirement for stakeholders to actively engage with the theory, which may not always be feasible.

- o Unique Contribution: The proposed framework simplifies the process of stakeholder engagement without compromising on depth. By focusing on straightforward, actionable steps like stakeholder classification, communication planning, and conflict resolution, the framework ensures that even complex stakeholder dynamics can be managed effectively without requiring extensive theoretical knowledge.

- Frameworks Focused on Specific Industries: Studies like those by Jayasena et al. [8] and Hassan et al. [9] focus on frameworks tailored to specific industries, such as smart city development or intelligent energy systems. These frameworks often offer valuable insights but may not be easily adaptable to other contexts due to their specialized nature.

- o Unique Contribution: The proposed framework is designed to be industry-agnostic, providing a flexible approach that can be adapted to various project types and industries. This versatility ensures that the framework can be broadly applied, offering a standardized yet customizable solution for diverse stakeholder management needs.

## 7. CONCLUSIONS AND FUTURE WORK

This study explores enhancing stakeholder engagement and requirements identification in project management by proposing a unified framework. The framework addresses challenges in stakeholder interactions by emphasizing early identification and continuous engagement throughout the project lifecycle. This structured approach ensures stakeholders' expectations are understood, aligning project goals with stakeholder needs to maintain project alignment and achieve successful outcomes. Validation of this solution was methodically carried out through an analysis model that included four specific goals, each designed to address a distinct aspect of stakeholder management. The effectiveness of the solution was measured against these goals, with data indicating strong support for the enhanced stakeholder identification and engagement processes proposed. Notably, high ratings in the areas of stakeholder communication and negotiation demonstrate the potential of the proposed framework to significantly improve project outcomes. Despite the promising results, the study also acknowledges areas for further improvement and refinement. The need for standardization in the stakeholder identification process and the adaptation of engagement strategies to accommodate varying project types and sizes were highlighted as potential areas for future development. Additionally, the findings suggest that further enhancements could include integrating advanced digital tools and providing more robust training for project managers to ensure the consistent application of these strategies across different projects.

Future work will focus on further refining the stakeholder engagement strategy to enhance its adaptability across different project types and sizes. This includes developing standardized processes for stakeholder identification and engagement to ensure consistency and reliability. Additionally, there will be an exploration of integrating advanced digital tools and technologies to streamline communication, data collection, and analysis. Future efforts will also emphasize creating comprehensive training programs for project managers to equip them with the necessary skills and knowledge to implement these strategies effectively. By addressing these areas, future work aims to optimize stakeholder management practices, ensuring better alignment with project objectives and improved project outcomes.

## AUTHORS

**Prof. Dr. Rizwan Qureshi** received his Ph.D. degree in Computer Sciences from National College of Business Administration & Economics, Pakistan 2009. He is currently working as a Professor in the Department of IT, King Abdulaziz University, Jeddah, Saudi Arabia. This author is the best researcher awardees from the Department of Information Technology, King Abdulaziz University in 2013 and 2016.

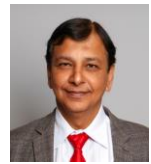

**Mr. Ahmed Abdulaziz Alnhari** is a master student in the Department of IT, King Abdulaziz University, Jeddah, Saudi Arabia.